\documentclass[a4paper]{revtex4}
\usepackage[english]{babel}
\usepackage{array,booktabs}% http://ctan.org/pkg/{array,booktabs}
\usepackage{array} 
\usepackage{lipsum}   
\usepackage{calc}
\usepackage{pdflscape}
\usepackage{color}
\usepackage{float}
\usepackage[font=small,labelfont=bf]{caption}
\usepackage{graphicx}
\usepackage{amsmath}
\usepackage[nodisplayskipstretch]{setspace}

\usepackage{setspace}
\usepackage{tabularx}

\usepackage{float}
\usepackage{color}
\usepackage{amsmath}
\usepackage{float}
\usepackage{calc}
\usepackage{pdflscape}
\usepackage{color}
\usepackage{float}
\usepackage{subfigure}
\usepackage[font=small,labelfont=bf]{caption}
\usepackage{graphicx}
\begin{document}{\setlength\abovedisplayskip{4pt}}

\title{Probing new physics in rare decays of b-flavored Hadrons $ b\rightarrow s \gamma $ in CMSSM/mSUGRA SUSY SO (10) theories.}

\author{Gayatri Ghosh}
\email{gayatrighsh@gmail.com}
\affiliation{Department of Physics, Gauhati University, Jalukbari, Assam-781015, India}

%\ead{{\color{blue}kalpana@gauhati.ac.in}}
%\ead{{\color{blue}gayatrighsh@gmail.com}}
%\ead{{e-mail: kalpana.bora@gmail.com}}
%\ead{{e-mail: gayatrighsh@gmail.com}}

\begin{abstract}
The implications of the latest measurement of the branching fraction of B($ b\rightarrow s \gamma $) of b hadrons, which is another signature of New Physics beyond Standard Model is presented here. The quark transitions $ b \rightarrow s $, $ b \rightarrow d $ do not happen at tree level in the Standard Model as the Z boson
does not couple to quarks of different flavour. In this work the present bounds on the quark transition $ b \rightarrow s $ within the constrained minimal supersymmetric extension of the Standard Model (CMSSM), in which there are three independent soft SUSY breaking parameters
$ m_{0} $, $ m_{1/2} $ and $ A_{0} $ is illustrated. The recent constraint on B($ b\rightarrow s \gamma $), B($ b_{s}\rightarrow \mu^{+}\mu^{-}$), the recently measured value of Higgs mass at LHC, $ M_{h} $, the value of $\theta_{13}$ from reactor data and the Higgs branching ratios  set very strong constraints on New Physics models, in particular supersymmetry. A new epoch for this research has begun since the Large Hadron Collider beauty (LHCb) experiment started affording data for various observables for these decays. The results presented here in mSUGRA/CMSSM models may gain access to supersymmetry even at scales beyond the direct reach of the LHC and the susceptibleness to test these theories at the next run of LHC is also explored.
\end{abstract}
\keywords{$ b\rightarrows  $ transitions, FCNCs,top quark, ILC, cMSSM/mSUGRA models.} 
\maketitle
\section{Introduction}	
\setlength{\baselineskip}{13pt}
Rare decays of hadrons containing a heavy bottom quark, denoted as
b hadrons, presents a powerful probe for exploring New Physics beyond Standard Model {\color{blue}\cite{a}}. b Hadrons decay most of the time via a $ b\rightarrow c W^{-*} $, where the W boson is virtual. These transitions are known as “tree decays” since the process binds a single mediator, the $ W^{-} $ boson. The quark transitions $b\rightarrow d \gamma  $, $b\rightarrow s \gamma  $ do not appear at tree level in the Standard Model as the
Z boson does not couple to quarks of different flavor. In the Standard Model the decay
$b\rightarrow s \gamma  $ occurs predominantly via a loop involving the top quark and the
W boson. It played a very vital role in flavour physics from the 1980s. It was the dependency of the branching fraction on the then unknown top quark mass that was the driving hot topic behind the theoretical calculations and the experimental searches. Flavor mixing in Standard Model quark sector is observed through processes like $ B^{0}-\bar{B^{0}} $ mixing and it came into view that the top quark was very heavy. The top quark was finally discovered at the Tevatron in 1995 and its mass was measured,
which predicted the Standard Model decay rate of $b\rightarrow s \gamma  $ to be around $ 10^{-4} $ {\color{blue}\cite{b}}. 
 \par 
Low energy observables implements interesting indirect information about the masses of supersymmetric particles. Predicting the masses of SUSY particles is more difficult
than for the top quark or even the Higgs boson mass, because the renormalizability of the Standard Model implicates that low-energy observables are impervious to heavy sparticles. Nonetheless present data on observables such as B($ b\rightarrow s \gamma $), B($ b_{s}\rightarrow \mu^{+}\mu^{-}$), the recently measured value of Higgs mass at LHC, $ m_{h} $, {\color{blue}\cite{Christoph}}, the value of $\theta_{13}$ from reactor data {\color{blue}\cite{Forero}} and the value of $ tan\beta $,the ratioof the MSSM Higgs vacuum expectation values provide interesting information on the scale of supersymmetry (SUSY).        
\par
The flavor changing neutral currents (FCNCs) are electroweak decays which proceeds through penguin Feynman diagrams with final state involving real photon or pair of leptons. Such decays were observed by CLEO experiments, CLEO (at
Cornell, USA), where it observed decay $ B\rightarrow K^{*} \gamma $. This opened the search for the inclusive decay, $ b\rightarrow s \gamma $.  Branching
fractions for these decays are $ 10^{-5}  $ or less, making the rare decay of b Hadrons as excellent candidates for quests for new physics beyond SM, setting stringent limits on the masses of the SUSY particles. In this work, the decay $ b \rightarrow s + \gamma $ is only considered, as this is unbeatably constrained by experiments. Recent calculation of the measured branching ratio of BR($ b \rightarrow s + \gamma $) at NNLO for $ E_{\gamma} \geq $ 1.6 GeV of $ (3.36 \pm 0.23) \times 10^{-4} $ is used in this calculation. Such experimental searches and theoretical studies on quark flavor violation can help us constrain the new physics beyond Standard Model that could be present just above the electroweak scale, or within the reach of the next run of LHC. Hitherto no 5 $ \sigma $ deviation from the Standard Model has been manifested in B Physics constraints so they suggests lower bound models assimilating New Physics.
\par 
Flavour-changing-neutral-current (FCNC) processes that make a down (d,s,b) or up (u,c,t) type quark convert it into another quark of the same type but of a different flavour are forbidden at tree level. These processes are suppressed by the Glashow-Iliopoulos-Maiani mechanism {\color{blue}\cite{glashow1970}} and involves at least one off-diagonal element of the CKM matrix. This in turn fabricates the processes rare.  
\par 
The value of the Higgs mass as measured at LHC {\color{blue}\cite{Christoph}}  and global fit values of the reactor mixing
angle $ \theta_{13} $ as measured at Daya Bay, Reno {\color{blue}\cite{Forero}}  have been used in this work.
\par 
In this work studies on rare decay of b Hadrons ($ b  \rightarrow s  \gamma $) using the type I seesaw mechanism in SUSY SO(10) theories is carried out, and hence the sensitivity to test the observation of sparticles at the next run of LHC {\color{blue}\cite{Christoph, candela, F}}, is checked in cMMSM/mSUGRA. minimal supergravity model. 
\par 
Here in this work model CMSSM/mSUGRA, where the soft terms are assumed to be universal at the Grand Unification (GUT) scale, $M_{GUT} = 2 \times 10^{16}$ GeV is explored. At the weak scale, the soft masses are no longer universal due to the effects of the renormalization group (RG) running. The minimal supergravity model (mSUGRA) is a well provoking model {\color{blue}\cite{Chamseddine, R.Barbarei, L.J.Hall, HP}}; for a review, see {\color{blue}\cite{R.L, RArnowitt, Ellis, E. Dudas}}; for reviews of the minimal supersymmetric standard model, see {\color{blue}\cite{X Tata, S. Dawa}}. In mSUGRA, SUSY is broken in the hidden sector and is communicated to the visible sector MSSM fields via gravitational interactions. The generation of gaugino masses
{\color{blue}\cite{E.Creme, L.E, pn, RL, L.A}} in mSUGRA (N = 1 supergravity) involves two scales $-$ the spontaneous SUGRA breaking scale in the hidden sector through the singlet chiral superfield and the other one is the GUT breaking scale through the non-singlet chiral superfield {\color{blue}\cite{Chamseddine, R.Barbarei, L.J.Hall, HP, R.L, RArnowitt, Ellis, E. Dudas}}. In principle these two scales can be different. But in a minimalistic viewpoint, they are usually assumed to be the same {\color{blue}\cite{Chamseddine, R.Barbarei, L.J.Hall, HP, R.L, RArnowitt, Ellis, E. Dudas}}. This leads to a common mass $m_{0}$ for all the scalars, a common mass $M_{1/2}$ for all the gauginos and a common trilinear SUSY breaking term $A_{0}$ at
the GUT scale,  $M_{GUT}\simeq2\times10^{16}$  GeV. Much more comprehensive and elaborate work exploring light supersymmetry exists in the literature, by the MasterCode {\color{blue}\cite{mm}} and
the Gambit collaborations {\color{blue}\cite{Gambit}}.
\par
SUSY is broken by soft terms of type $- A_{0}, m_{0}, M_{1/2}$, where $A_{0}$ is the universal trilinear coupling, $m_{0}$ is the universal scalar mass, and $M_{1/2}$ is the universal gaugino mass. Stern universality between Higgs and matter fields of mSUGRA models can be relaxed in NUHM {\color{blue}\cite{EllisJ}} models.
\par 

In the case of tan $ \beta  $ = 50, in order to have Higgs mass $ m_{h} $ around 125.09 GeV, values of $M_{1/2} \leq $ 1000 GeV is strongly disfavoured. As shown in our results in Sect. {\color{blue}III} in mSUGRA, the spectrum of $M_{1/2}$ and $ m_{0} $ is found to lie toward the heavy side, as allowed by recent constraints on BR($ b \rightarrow s \gamma $), for $ tan \beta  = 50$ as compared to the case for $ tan \beta  = 10$ where only lighter spectra are possible. It is seen in this work that in $M_{1/2}-tan \beta$ plot B
physics constraints prefer low mass of $ M_{1/2} $ particles. It is owing to the reality
that the MSSM corrections to the B physics observables inversely depend on the charged
Higgs particles masses, $ H^{\pm} $.
 
\par 
From above it is seen that the indication of New Physics beyond Standard Model could be tested at the next run of LHC, if the SUSY sparticles are observed within a few TeV. No SUSY partner of SM has been been observed yet, and this could point to a high scale SUSY theory. The LHC has stringent consytaints on the sparticles, which indicates a tuning of EW symmetry at a few percent level {\color{blue}\cite{TG, A.Arv, EHard, JL, T. Gher, Fan}}.

\par
In this paper it is shown how studies of rare decays of B mesons probe any Physics beyond Standard Model. The rare decay $ b \rightarrow s + \gamma $ is investigated in CMSSM model in the light of LHC bounds and investigate whether present or future run of LHC can access Supersymmetry. The rest of the paper is organized as follows. In section {\color{blue}II}, the effective Hamiltonian and the observables in $ b \rightarrow s + \gamma $ that are sensitive to the scale of supersymmetry is discussed. In section {\color{blue}III} a numerical analysis of the current constraints on new physics in the $ b \rightarrow s + \gamma $ transition is discussed, taking into account the theoretical uncertainties. Section {\color{blue}IV}
summarises our conclusions.
\section{Rare Decay $ b \rightarrow s \gamma $ decay in mSUGRA/CMSSM model of SUSY SO(10) theory}
It is admitted that the experimental bounds of $ b \rightarrow s \gamma $ decay set very stringent constraints on CMSSM/mSUGRA model, minimal extension of the Stan
dard Model (SM) {\color{blue}\cite{c}}. The $ \tan \beta $ enhanced radiative corrections, which springs from the renormalization of the Yukawa coupling to down-type fermions, can productively be summed to all orders in perturbation theory and can be formed into an effective lagrangian {\color{blue}\cite{d}} :
\begin{equation}
L = -h^{ij}_{d}\bar{d}^{i}_{R}H_{1}Q^{j}_{L}-\delta h^{ij}_{d}\bar{d}^{i}_{R}H_{2}Q^{j}_{L}-\delta h^{ij}_{u}\bar{u}^{i}_{R}(i\tau_{2}H^{*}_{2})Q^{j}_{L}-\delta h^{ij}_{u}\bar{u}^{i}_{R}(i\tau_{2}H^{*}_{1})Q^{j}_{L}+h.c.,
\end{equation}
where $ \tau_{2} $ is the two by two Pauli matrix, $ Q_{L} = (u, d)_{L} $, and a gauge-invariant contraction of weak and colour indices has been totally presumed. 
Inferring that the right and left handed soft-supersymmetry-breaking
mass parameters are generation-independent, they are proportional to the couplings $ h_{d} $ and $ h_{u} $, 
\begin{equation}
\delta h^{SQCD}_{d} = \frac{2\alpha_{s}}{3\pi}M_{\tilde{g}}\mu I(m_{\tilde{b_{L}}}, m_{\tilde{b_{R}}},M_{\tilde{g}}),
\end{equation}
\begin{equation}
\delta h^{SQCD}_{u} = \frac{2\alpha_{s}}{3\pi}M_{\tilde{g}}\mu I(m_{\tilde{t_{L}}}, m_{\tilde{t_{R}}},M_{\tilde{g}})
\end{equation}
where $M_{\tilde{g}} $ is the gluino mass and $ m_{\tilde{b_{L,R}}} $, $ m_{\tilde{t_{L,R}}} $  are the left and right-handed mass parameters of the down- and up-squarks respectively.
\par
where, I(a,b,c) is defined in {\color{blue}\cite{d}}. In application to $ b \rightarrow s \gamma $
one requires the coupling of the charged Higgs boson to the right-handed bottom and left-handed top quarks, for which, overlooking small CKM angle effects, a large
tan $ \beta $-enhanced corrections is achieved by replacing the tree-level connection between the coupling $ h_{b} $ in {\color{blue}Eq.1} and the bottom mass by
\begin{equation}
h_{b} =  \frac{g}{1.414 M_{W} cos\beta}\frac{\bar{m_{b}(Q)}}{1+\Delta m_{b}^{SQCD}},
\end{equation} 
Where g is the SU(2) gauge coupling, $ M_{W} $ is the mass of W boson, and $ m_{b} $ is the bottom mass renormalized in a mass-independent renormalization scheme. The $ tbH^{+} $ vertex is renormalized at the scale Q, which gets into $ m_{b} $ in {\color{blue}Eq.4} . When claimed into $ b \rightarrow s \gamma $  the scale Q equals the scale $ \mu_{W} $, at which top and charged Higgs Boson $H^{+}$ are integrated out. The $ tan\beta $ SQCD corrections are;
\begin{equation}
\Delta m^{SQCD}_{b} = \frac{2 \alpha_{s}}{3\pi}M_{\tilde{g}}\mu tan\beta I(m_{\tilde{b_{L}}}, m_{\tilde{b_{R}}},M_{\tilde{g}}),
\end{equation}
where $ \alpha_{s} $ is evaluated at a scale of the order, where masses enter I.
The $ tan \beta $ enhanced chargino benefactions to $ b \rightarrow s \gamma $ are, 
\begin{equation}
BR|(b \rightarrow s \gamma)|_{\chi^{\pm}} \propto \mu A_{t}tan\beta f(m_{\tilde{t1}},m_{\tilde{t2}},m_{\tilde{\chi}^{+}})\frac{m_{b}}{v(1+\Delta m_{b})},
\end{equation}
All presiding higher-order contributions are embraced in $ \Delta m_{b} $, and f is the loop integral occuring at one loop. In large tan$ \beta $ region, the admissible charged Higgs contribution to BR$(b \rightarrow s \gamma)$ are,
\begin{equation}
BR|(b \rightarrow s \gamma)|_{H^{+}} \propto \frac{m_{b}(h_{t}cos\beta-\delta h_{t}sin\beta)}{vcos\beta(1+\Delta m_{b})}g(m_{t},m_{H^{+}}),
\end{equation}
g is the loop-integral rising at the one-loop level. In the above $ \delta h_{t} $ results from the flavour violating coupling $ \delta h_{u} $  in {\color{blue}Eq.1}.
\par 
At large values of $ tan \beta $, the leading order chargino subscriptions to the amplitude of $ b \rightarrow s \gamma $ are proportional to $ \mu A_{t} $ or $ \mu A_{0} $. In mSUGRA models, the sign of $ A_{t} $ is opposite to the one of the gaugino masses. This sign relation clinch except for the boundary values of $ A_{t} $ at some
high-energy input scale is one order of magnitude bigger than the  soft susy breaking mass parameters of gaugino masses {\color{blue}\cite{e,f}}. It is shown in {\color{blue}\cite{d}} how negative values of $ \mu $ the next to leading-order corrections to the charged Higgs and the chargino-stop contributions further enlarges the $ b \rightarrow s \gamma $ decay amplitude. Even after considering higher order effects, positive values of $ \mu $ render it necessary to access correct values for BR($ b \rightarrow s \gamma $)  within CMSSM/mSUGRA models, for which the sign of $ A_{t} $ or $ A_{0} $ at low energies leans to be negative, which is precisely shown in this work.
 It may be noted that some results on neutrino masses and mixings using updated values of running quark and lepton masses in SUSY SO(10) is discussed in {\color{blue}\cite{GGhosh}}, which leads to lepton flavor violation in SUSY SO(10) models {\color{blue}\cite{CGhosh}}. Also some studies on non unitarity of PMNS matrix in the leptonic sector has been done in {\color{blue}\cite{P,Q}}
\section{Calculation of BR($b \rightarrow s\gamma$) in mSUGRA}
In this section calculations and results on the
rare decay of b Hadrons in CMSSM SUSY SO(10) theory is presented, with the type I seesaw mechanism using the mSUGRA boundary conditions through detailed numerical analysis. The soft parameter space for CMSSM/mSUGRA is scanned in the following ranges :
$$ tan\beta=10 $$
$$ tan\beta=50 $$
$$ m_{h} \in \left[ 122.5, 129.5\right] \hspace{.1cm}\text{GeV}$$
$$ \Delta m_{H} \in 0 $$
$$ m_{0} \in \left[ 0, 5\right] \hspace{.1cm}\text{TeV} $$
$$ M_{1/2} \in \left[ 0, 2\right] \hspace{.1cm} \text{TeV}$$, 
The choice of $ A_{0} $ used in this analysis are;
$$ A_{0}  = 0\hspace{.1cm} \text{TeV} $$
$$ A_{0} \in \left[ -M_{1/2} , + M_{1/2} \right]$$
$$ A_{0} \in \left[ -2M_{1/2} , + 2M_{1/2} \right]$$
$$ A_{0} \in \left[ -3M_{1/2} , + 3M_{1/2} \right]$$
\begin{equation}
sgn\left( \mu\right) \in\lbrace-,+\rbrace, 
\end{equation}
\vspace{.05cm}
The numerical analysis is carried out using the publicly available SUSEFLAV package {\color{blue}\cite{Garani}}. The program calculates $BR(b \rightarrow s+\gamma)$ in the minimal flavor violation assumptions. Here $ m_{0} $ is the universal soft SUSY breaking mass parameter for sfermions, and $ M_{1/2} $ denote the common gaugino masses for $U(1)_{Y}$, $SU(2)_{L}$ and $SU(3)_{C}$. $ A_{0} $ is the trilinear scalar interaction coupling, $ \text{tan}\beta $  is the ratio of the MSSM Higgs vacuum expectation values (VEVs).
\par
The masses of the heavy neutrinos used in this calculations are - $ M_{R_{1}}=10^{13} \hspace{.1cm}\text{GeV}$, $M_{R_{2}} = 10^{14}\hspace{.1cm}\text{GeV}$, and $M_{R_{3}}=  10^{16}\hspace{.1cm}\text{GeV}$. For $ \Delta m^{2}_{sol} $, $\Delta m^{2}_{atm}$ and $\theta_{13}$, the central values from the recent global fit of neutrino data {\color{blue}\cite{Forero}} are used. 
\section{Analysis and discussion of results}
In this section, analysis and discussion of results obtained in Sect. {\color{blue}III} in Complete universality: cMSSM (mSUGRA) is presented. Also the sensitivities to the scale of supersymmetry from low-energy observables used in this analysis is also discussed.
\subsection{ The rare decay of b Hadrons $BR(b \rightarrow s+\gamma)$   }
In constrained MSSM (CMSSM) model, the soft supersymmetry-breaking mass parameters, scalar masses and gaugino masses are each assumed to be equal at GUT scale, $ M_{GUT} $. In this case, the new independent MSSM parameters are four in numbers.
In mSUGRA at the high scale, the parameters of the model are the scalar mass $m_{0}$, the trilinear soft supersymmetry-breaking parameter $A_{0}$, and the unified gaugino mass $ M_{1/2} $. Besides, there is the Higgs potential parameter $ \mu $ and the undetermined ratio of the Higgs VEVs, tan$\beta$. The whole supersymmetric mass spectrum is decided once these parameters are set. The updated measured branching fraction $ BR(b \rightarrow s \gamma) $ {\color{blue}\cite{b}} together with a large $\theta_{13}$ {\color{blue}\cite{Forero}} places striking constraints on the SUSY parameter space in CMSSM. In Fig. {\color{blue}1a$-$1d} and Fig. {\color{blue}2a$-$2d} the constraints from  $BR(b \rightarrow s+\gamma)$ on mSUGRA parameter space for tan$ \beta $ = 10 and tan $ \beta $ = 50 is presented. Each Fig. {\color{blue}1a,1c}, Fig. {\color{blue}2a,2c}, in the left panel displays constrained allowed parameter region for tan$ \beta $ = 10, the Fig. {\color{blue}1b,1d}, Fig. {\color{blue}2b,2d}, in the right panel dispose preferred region for tan$ \beta $ = 50. As can be seen from Fig. {\color{blue}1}a, almost no part of the paramater space survives for tan $ \beta $ = 10, but a significant amount of parameter space $M_{1/2}\sim 800  $ GeV to $M_{1/2}\sim 1500  $ GeV exists for tan $ \beta $ = 50 for $ A_{0}  = 0\hspace{.1cm} \text{TeV} $ in mSUGRA as allowed by the updated BR($ b \rightarrow s \gamma = 3.36 \pm 10^{-4}$ ) as is evident from Fig.{\color{blue}1}c. Also allowed parameter  space from almost $M_{1/2}\sim 400$ GeV to $M_{1/2}\sim 750$ GeV is permitted in reduced cluster as is perceivable from Fig.{\color{blue}1}b. This leads to the conclusion that the parameter space $M_{1/2} \geq $ 1.5 TeV is not allowed by the present upper bounds on BR($ b \rightarrow s \gamma < 3.59 \times 10^{-4}$). The allowed regions in Fig. {\color{blue}1}c require very light spectra, i.e. $ M_{1/2}$ lies in between 400 GeV to 750 GeV for tan$ \beta $ = 10, $ A_{0} \in \left[ -M_{1/2} , + M_{1/2} \right]$, as allowed by the present measured branching fraction $ B(b \rightarrow s \gamma) $ bounds. In Fig. {\color{blue}1}d the whole parameter space from $M_{1/2}\sim 750 $ GeV to $M_{1/2}\sim 2$ TeV is acceptable for tan$ \beta $ = 50, $ A_{0} \in \left[ -M_{1/2} , + M_{1/2} \right]$. Also dispersed region for $M_{1/2}$ lying betwen 400 GeV to 700 GeV is admissible in in Fig. {\color{blue}2}a. Next, the results obtained in CMSSM case for $ A_{0} \in \left[ -2M_{1/2} , + 2M_{1/2} \right]$, $ A_{0} \in \left[ -3M_{1/2} , + 3M_{1/2} \right]$ is introduced. In Fig. {\color{blue}2}a $M_{1/2}$ vs. log[BR($ b \rightarrow s+\gamma$)is conferred for tan$ \beta $ = 10, $ A_{0} \in \left[ -2M_{1/2} , + 2M_{1/2} \right]$ and the Fig. {\color{blue}2}b in the right panel shows $ M_{1/2} $[GeV] vs. log[BR($ b \rightarrow s+\gamma$) for tan$ \beta $ = 50, $ A_{0} \in \left[ -2M_{1/2} , + 2M_{1/2} \right]$. Different horizontal lines in Fig. {\color{blue}2a,b,c,d} correlates with present bounds on BR($ b$ $ \rightarrow $ s + $ \gamma $). It is found from  Fig. {\color{blue}2a}, very few parameter space survises for, $ tan\beta= $ 10. $ M_{1/2} $ almost lies in the range, 400 GeV$\sim$ 700 GeV.  It is seen  from Fig. {\color{blue}2}b
that most of the mSUGRA parameter space is going to be explored by present bounds on the measured branching ratio, BR($ b$ $ \rightarrow $ s + $ \gamma $) at NNLO for $ E_{\gamma} \geq $ 1.6 GeV of $ (3.36 \pm 0.23) \times 10^{-4} $ {\color{blue}\cite{b}}. 
\par 
In Fig.{\color{blue}2c,2d}, the SUSY parameter space  of $M_{1/2} $ is presented, as allowed by present bounds on BR($ b$ $ \rightarrow $ s + $ \gamma $), at NNLO for $ E_{\gamma} \geq $ 1.6 GeV i.e, $ (3.36 \pm 0.23) \times 10^{-4} $, $ A_{0} \in \left[ -3M_{1/2} , + 3M_{1/2} \right]$ .
For smaller tan $\beta  =$ 10, even values of $ M_{1/2} $ as low as $ \sim $ 400 GeV would be allowed if tan $\beta = $ 10, whereas $ M_{1/2}\geq $ 1 TeV would be required if tan $\beta = $ 50. These limits are very sensitive to $ A_{0} $. Thus the combination
of $ B(b \rightarrow s + \gamma) $ together with the other precision observables might be able, in principle, to constrain$ A_{0} $ strikingly. It is found that a wide region of soft parameter space for tan $ \beta = $ 50 is allowed  which
would be easily attainable at the next run of LHC satisfying the current $ B(b \rightarrow s + \gamma) $ constraints. 
\begin{center}
\begin{figure*}[htbp]
\centering{
\begin{subfigure}[]{\includegraphics[height=6.9cm,width=7.9cm]{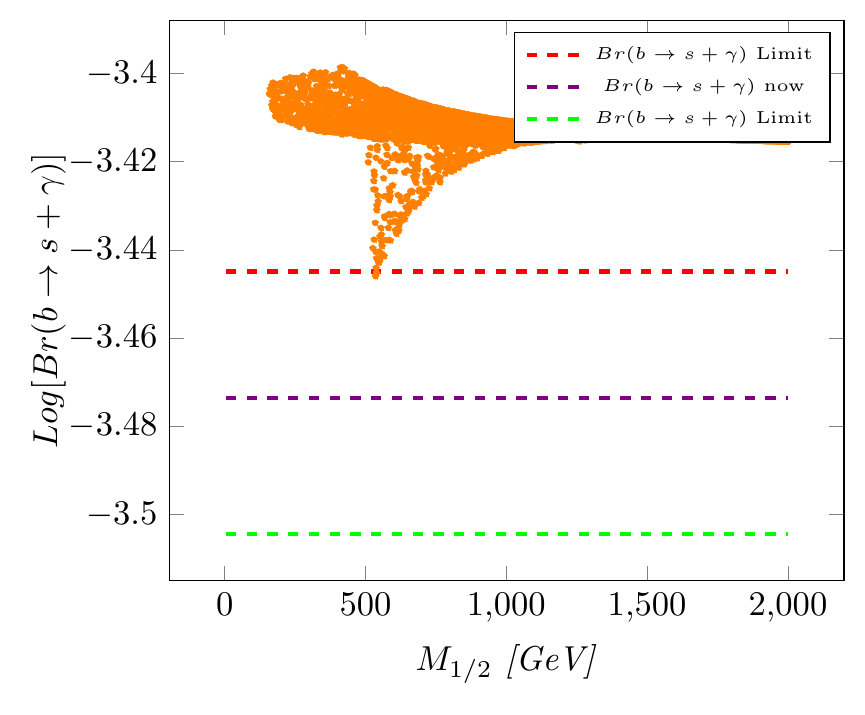}}\end{subfigure}
\begin{subfigure}[]{\includegraphics[height=	6.93cm,width=8cm]{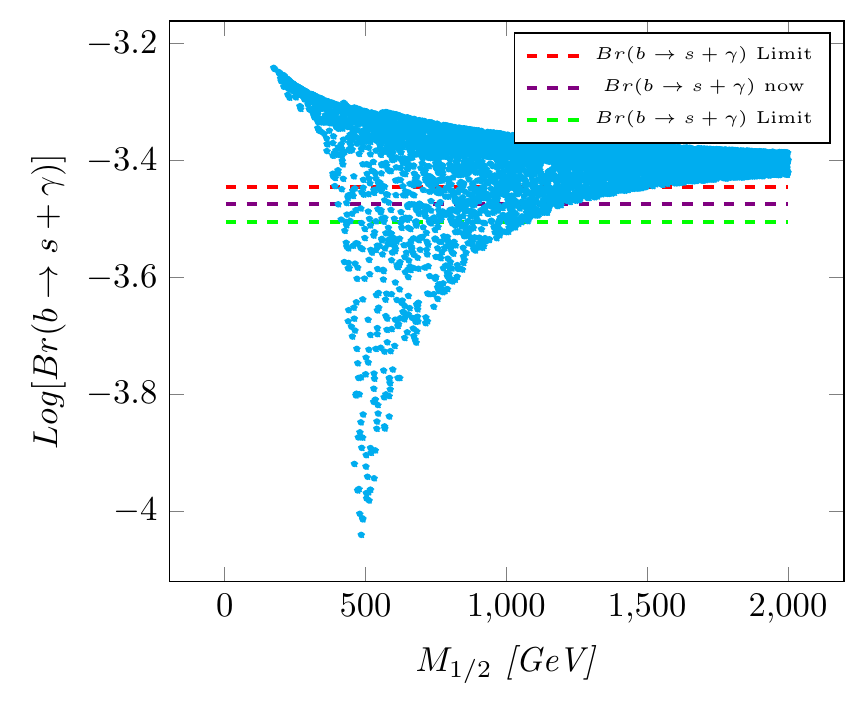}}\end{subfigure}\\
\begin{subfigure}[]{\includegraphics[height=6.9cm,width=7.9cm]{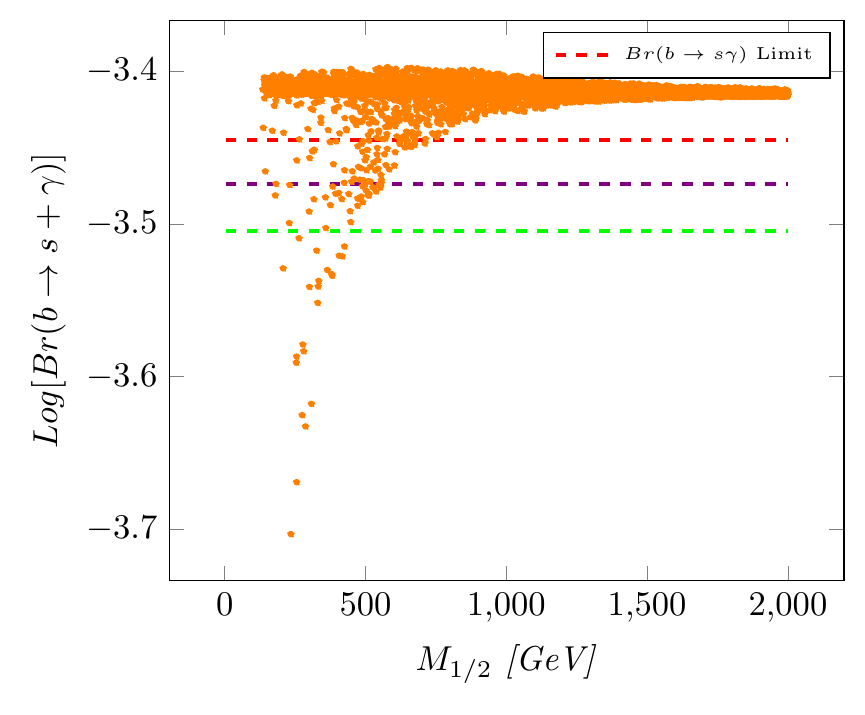}}\end{subfigure}
\begin{subfigure}[]{\includegraphics[height=6.9cm,width=8cm]{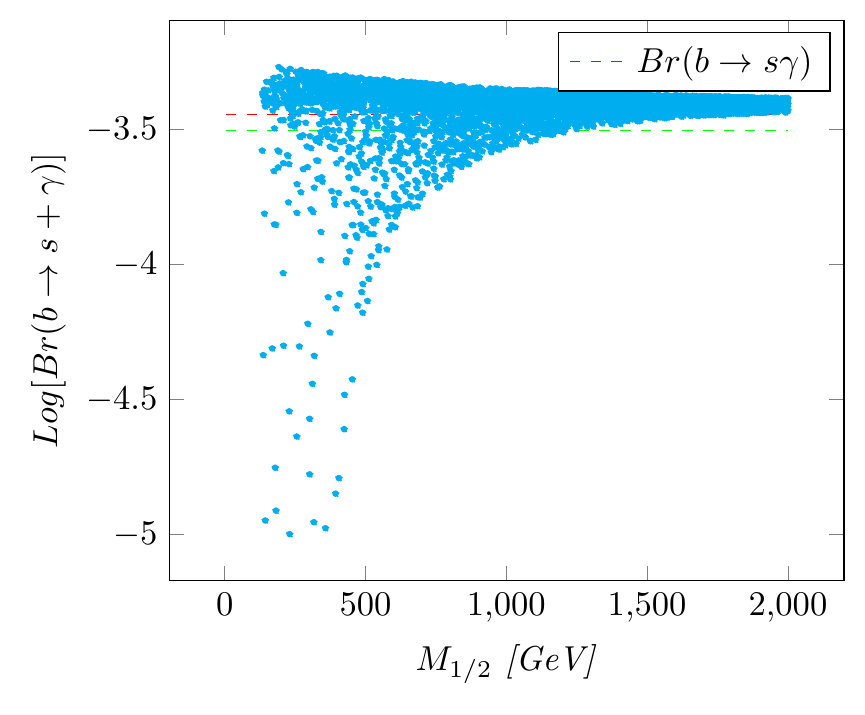}}\end{subfigure}\\
\caption{The CMSSM predictions for BR($ b $ $ \rightarrow $ s
 + $ \gamma $) as a function of $ M_{1/2} $ (a) tan $ \beta = 10 $ in the left panel and (b) tan $ \beta = 50 $ in the right panel and for various choices of $ A_{0} $. Figures in the top panel is for $ A_{0}  = 0\hspace{.1cm} \text{TeV} $, and the figures in the bottom panel consider $ A_{0} $ to be lying in the range, $ A_{0} \in \left[ -M_{1/2} , + M_{1/2} \right]$. The central (dotted) violet line indicates the current experimental central value of BR($ b $ $ \rightarrow $ s
 + $ \gamma $),
and the other dotted orange and green line splashes its current value within its $ \pm $ 1 $ \sigma $ ranges.}}
\end{figure*}
\end{center}

\begin{center}
\begin{figure*}[htbp]
\centering{
\begin{subfigure}[]{\includegraphics[height=6.9cm,width=7.7cm]{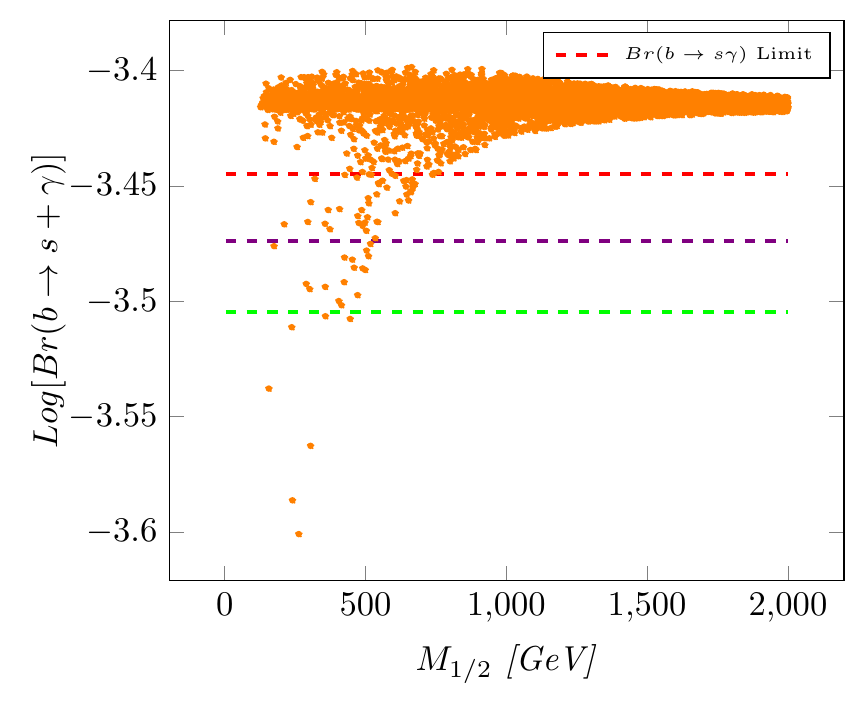}}\end{subfigure}
\begin{subfigure}[]{\includegraphics[height=6.9cm,width=7.9cm]{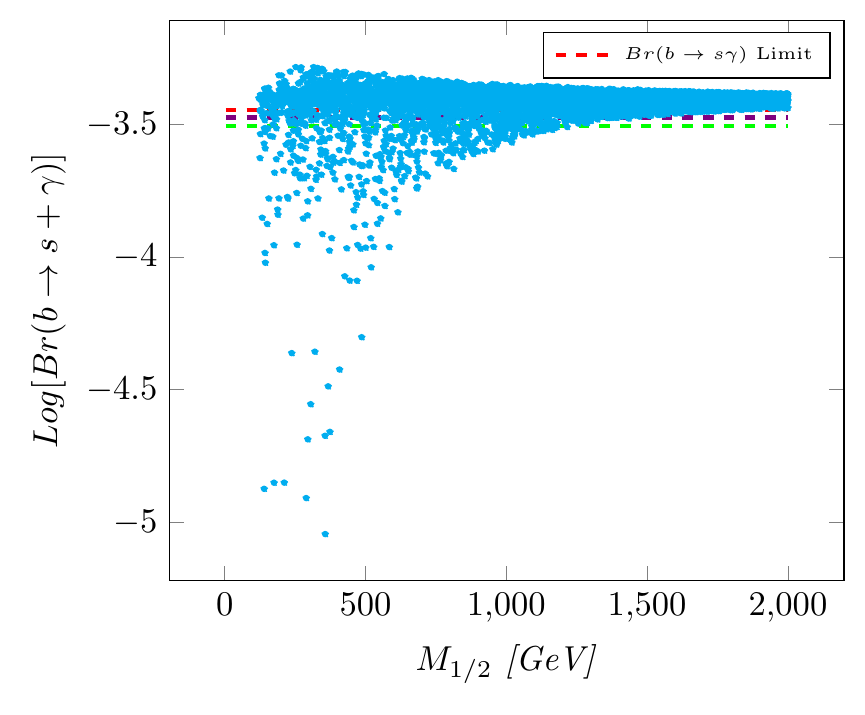}}\end{subfigure}\\
\begin{subfigure}[]{\includegraphics[height=6.9cm,width=7.7cm]{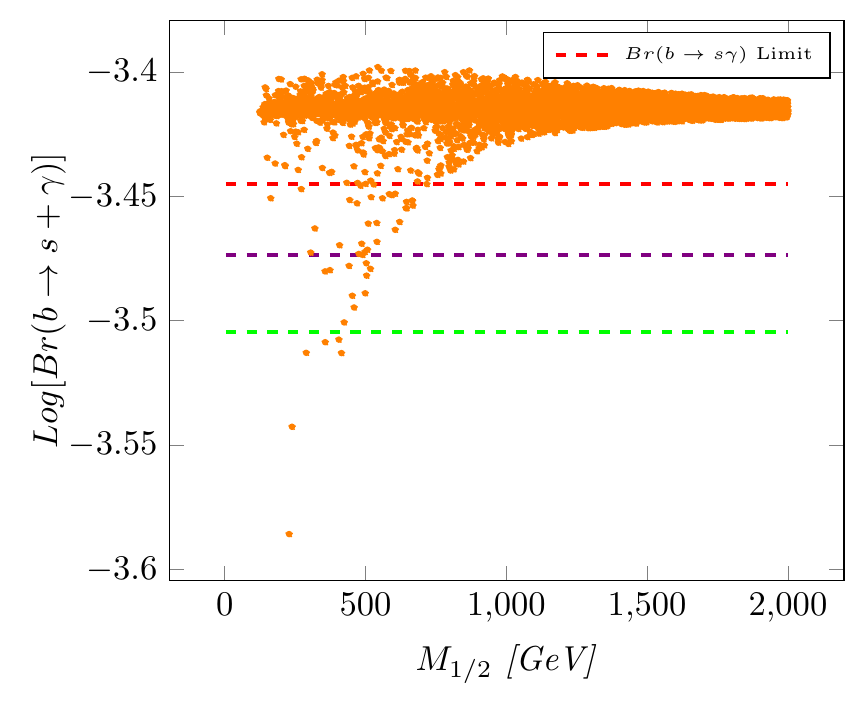}}\end{subfigure}
\begin{subfigure}[]{\includegraphics[height=6.9cm,width=7.9cm]{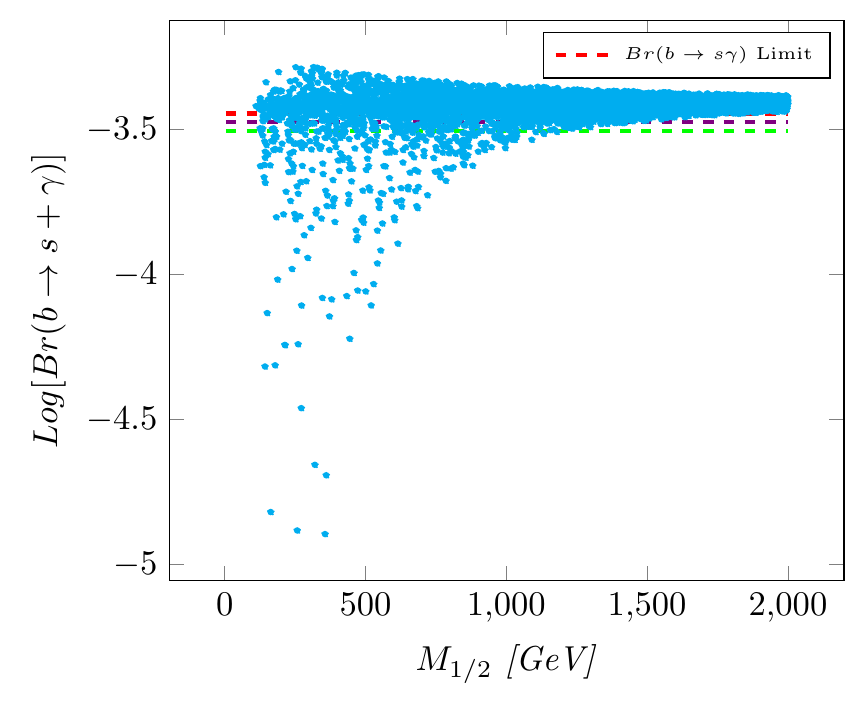}}\end{subfigure}\\
\caption{The CMSSM predictions for BR($ b $ $ \rightarrow $ s
 + $ \gamma $) as a function of $ M_{1/2} $ (a) tan $ \beta = 10 $ in the left panel and (b) tan $ \beta = 50 $ in the right panel and for various choices of $ A_{0} $. Figures in the top panel is for $ A_{0} \in \left[ -2M_{1/2} , + 2M_{1/2} \right]$, and the figures in the bottom panel consider $ A_{0} $ to be lying in the range, $ A_{0} \in \left[ -3M_{1/2} , + 3M_{1/2} \right]$. The central (dotted) violet line indicates the current experimental central value of BR($ b $ $ \rightarrow $ s
 + $ \gamma $),
and the other dotted orange and green line splashes its current value within $ \pm $ 1 $ \sigma $ ranges.}}
\end{figure*}
\end{center}
\subsection{\textbf{The Lightest neutral MSSM higgs boson mass}}
 The discovery in 2012 by the ATLAS and the CMS collaborations of the Higgs boson mass was a significant turning point as well as a spectacular success of the ATLAS and CMS experiments. The ATLAS and CMS experiments have made united measurement of the mass of the Higgs boson in the diphoton and the four photon channels at per mille accuracy, $ m_{h} = 125.09 \pm 0.24 GeV$.
 \par 
It is shown in Fig.{\color{blue}3a$-$3f} the high sensitivity of the electroweak precision observable, Higgs boson mass i.e $ m_{h} = 125.09 \pm 0.24 GeV$  to constrain the supersymmetric
parameter space. In Fig. {\color{blue}3a, 3c, 3e (3b, 3e, 3f)} the constraint from  $ m_{h} = 125.09 \pm 0.24 GeV$ on CMSSM parameter space for tan$ \beta $ = 10 (tan$ \beta $ = 50) is presented. Different horizontal lines displays the ATLAS and CMS results of the  bounds on Higgs boson mass with uncertainity, $ m_{h} = 125.09 \pm 0.24 GeV$ within its 1$ \sigma $ values. It is seen from Fig. {\color{blue}3a, 3c} no part of the mSUGRA parameter space convinces the restrictions set by MSSM lightest CP even Higgs boson mass. For, $ A_{0} \in \left[ -2M_{1/2} , + 2M_{1/2} \right]$ it is established from Fig. {\color{blue}3}d gaugino masses $M_{1/2} \geq $ 1.5 TeV corresponding to 125.09 GeV Higgs are mostly disfavored, which would be congenial at the future run of LHC, satisfying the current bounds on $BR( b \rightarrow s+\gamma) \simeq (3.36 \pm 0.23) \times 10^{-4})$. From {\color{blue}Fig. 3f}, for which $ A_{0} \in \left[ -3M_{1/2} , + 3M_{1/2} \right]$, it is seen that for a Higgs mass around $ m_{h} = 125.09 \pm 0.24 GeV$, $ M_{1} $  with its uncertainity, $ M_{1/2} $ lies between
 $ 1.1 \hspace{.1cm}\text{TeV} \leq M_{1}\leq 2 \hspace{.1cm} \text{TeV}$. The constraints imposed on the soft SUSY breaking parameters of CMSSM, for tan$ \beta $ = 50,  are found
to be less severe compared to tan$ \beta $ = 10 for which, no part of the parameter space satisfy the constraints put by the measured lightest neutral Higgs mass at LHC. The contrast between the measured value of $ m_{h} = 125.09 \pm 0.24 GeV$ and an exact theory prediction
will entitle one to set stiff constraints on the allowed parameter space
of $ M_{1/2} $.

\begin{center}
\begin{figure*}[htbp]
\centering{
\begin{subfigure}[]{\includegraphics[height=5.9cm,width=7.9cm]{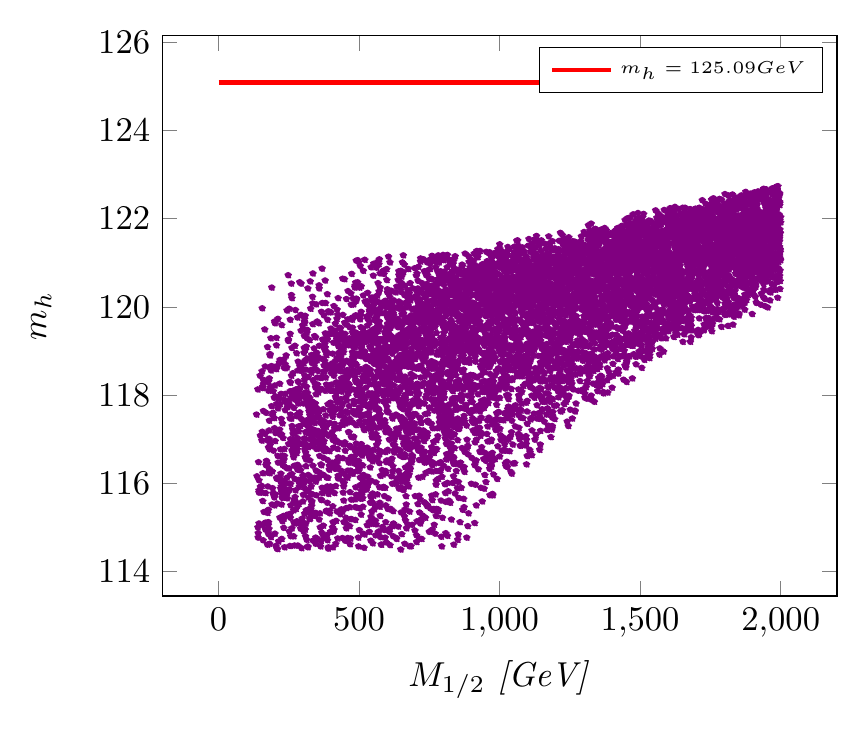}}\end{subfigure}
\begin{subfigure}[]{\includegraphics[height=5.9cm,width=8cm]{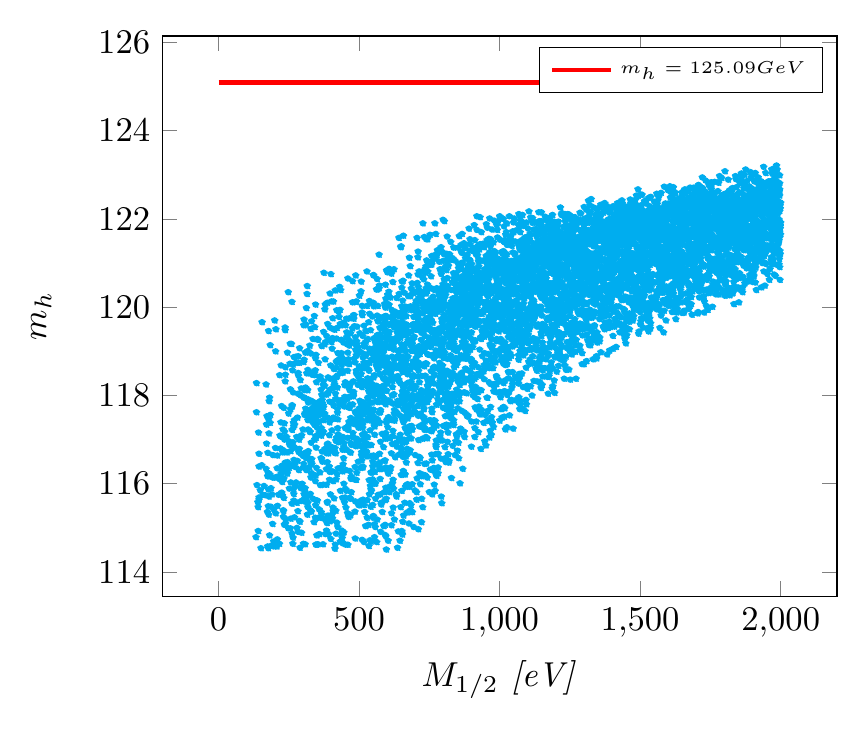}}\end{subfigure}\\
\begin{subfigure}[]{\includegraphics[height=7.9cm,width=7.9cm]{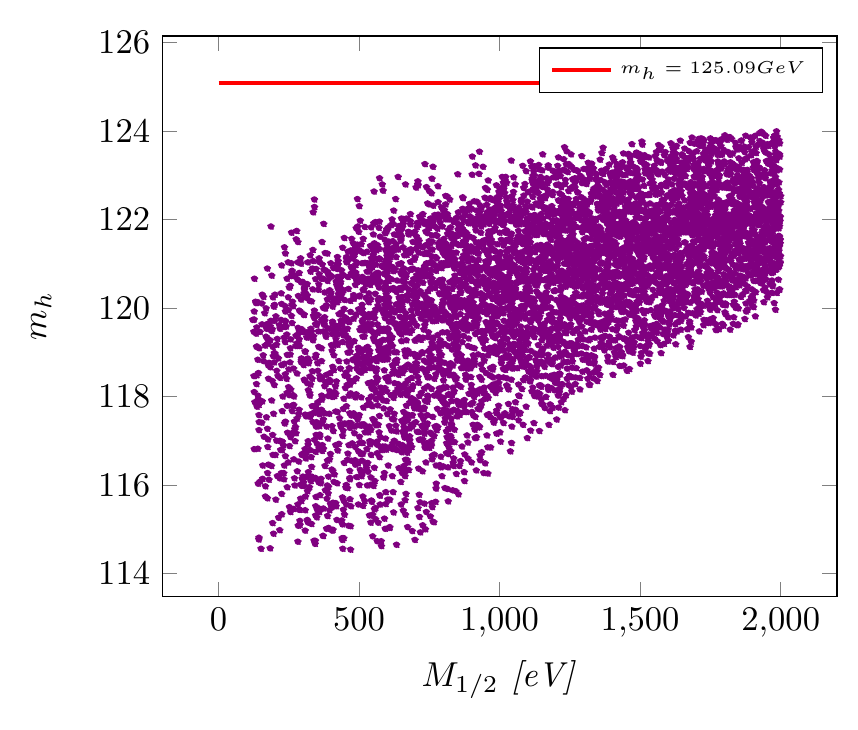}}\end{subfigure}
\begin{subfigure}[]{\includegraphics[height=7.9cm,width=8cm]{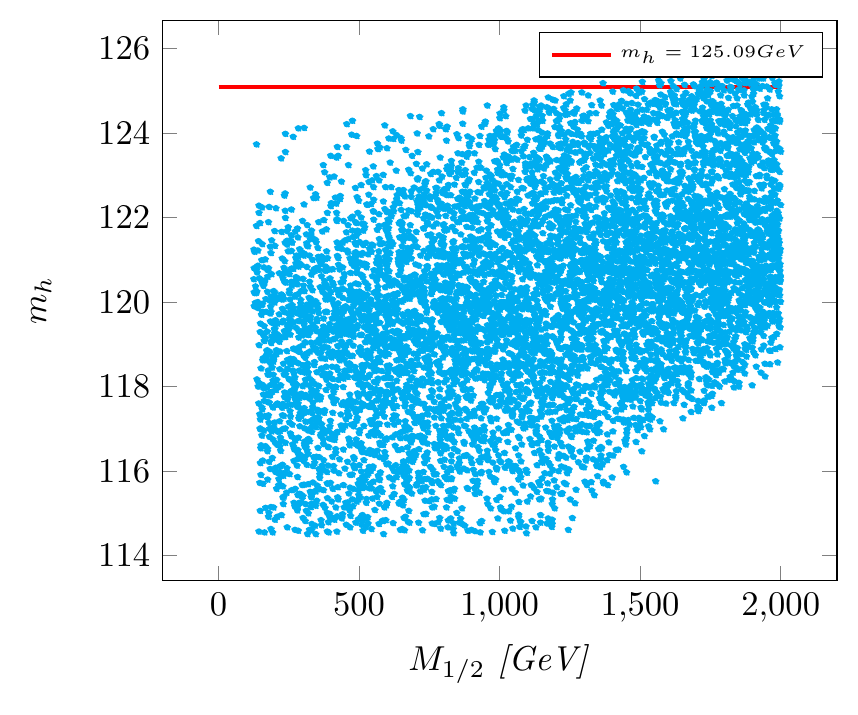}}\end{subfigure}\\
\begin{subfigure}[]{\includegraphics[height=7.9cm,width=7.9cm]{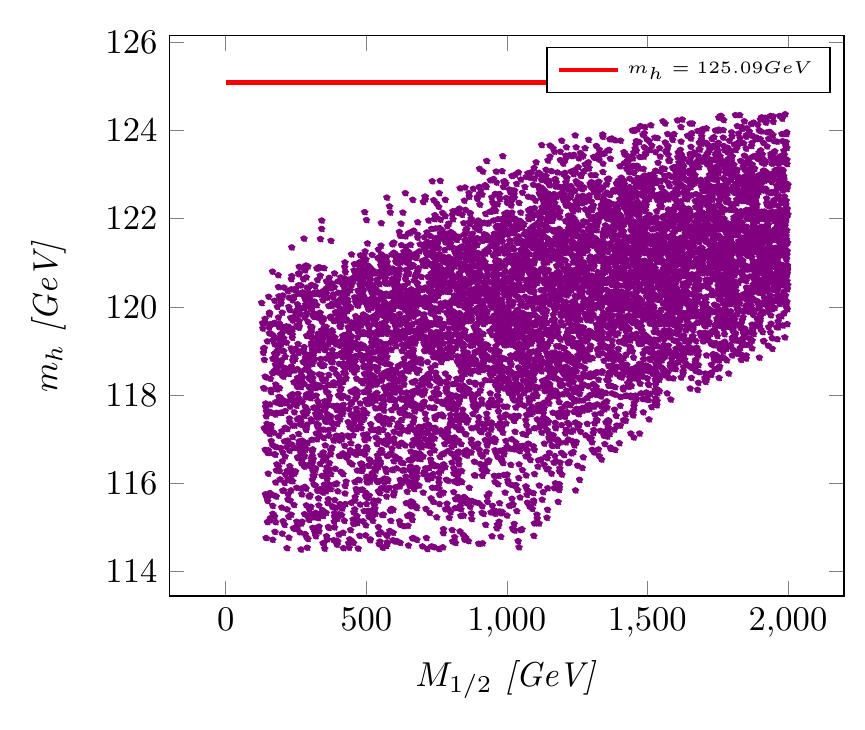}}\end{subfigure}
\begin{subfigure}[]{\includegraphics[height=7.9cm,width=8cm]{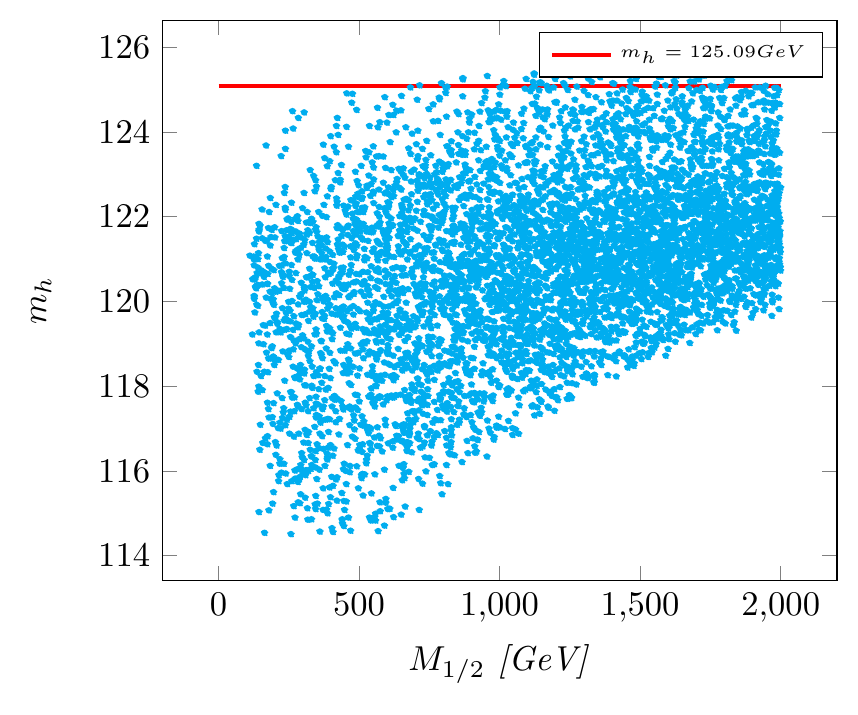}}\end{subfigure}\\

\caption{The CMSSM predictions for $m_{h}$ as a function of $ M_{1/2} $ constrained by recent experimental bounds on BR($ b \rightarrow s+
\gamma$) for (a) tan $ \beta = 10 $ in the left panel and (b) tan $ \beta = 50 $ in the right panel and for various choices of $ A_{0} $. Figures in the upper panel is for $ A_{0} \in \left[ -M_{1/2} , + M_{1/2} \right]$ , and the figures in the second panel consider $ A_{0} $ to be lying in the range, $ A_{0} \in \left[ - 2 M_{1/2} , + 2 M_{1/2} \right]$. Figures in the third panel represent $ A_{0} $ to be lying in the range, $ A_{0} \in \left[ -3 M_{1/2} , + 3 M_{1/2} \right]$. The solid red line indicates the current experimental central value of $ m_{h} $ from the ATLAS and the CMS collaboration, $ m_{h} = 125.09  $ GeV.}}
\end{figure*}
\end{center}
\subsection{\textbf{Scan of the mSUGRA parameter space}}

From the studies in mSUGRA model in the above subsections, we see that the SUSY parameter space, as allowed by bounds on BR($b \rightarrow s+ \gamma$)
shifts to the lighter side for $ tan \beta = 50 $. The
sensitivity of the precision observables $BR( b \rightarrow s+\gamma) \sim (3.36 \pm 0.23) \times 10^{-4}$ in a scan over the $(M_{1/2}, A_{0})$, $(M_{1/2}, m_{0})$ parameter plane is studied.  Figure {\color{blue}4}a(b) shows $ m_{0} $ [GeV] Vs $ M_{1/2} $ [GeV] as allowed by the present bound on BR$(b \rightarrow s \gamma)$ for $ tan \beta = 50 $, $ A_{0} \in \left[ -2M_{1/2} , + 2M_{1/2} \right]$, $ A_{0} \in \left[ -3M_{1/2} , + 3M_{1/2} \right]$. We find that the $ M_{1/2} \geq $ 100 GeV and $ m_{0} \geq $ 0.5 TeV region is mostly favoured. 
\par 
The fit for the allowed region of ($M_{1/2}, A_{0}$) plane is depicted in Figure {\color{blue}5}a,c (b,d) for $ tan \beta =10 $ ($ tan \beta = 50 $) and $ A_{0} \in \left[ -2M_{1/2} , + 2M_{1/2} \right]$ , $ A_{0} \in \left[ -3M_{1/2} , + 3M_{1/2} \right]$ respectively.  
It is grasped from Figure {\color{blue}5}b, that the permitted parameter space constrained by the measured branching ratio $BR( b \rightarrow s+\gamma) \sim (3.36 \pm 0.23) \times 10^{-4}$ is obtained for $ A_{0} \leq $ 0, while positive values of
$ A_{0}$ result in a somewhat lower fit quality. The scan yields an upper bound on
$ M_{1/2} $ of about 1.8 TeV.  From Fig. {\color{blue}5d} it is seen that the allowed spectrum is for negative $ A_{0} $. The upper bound on
$ M_{1/2} $ increases to nearly 1.65 TeV. A slightly higher sensitivity for
$ A_{0} \leq 0 $ than for positive $ A_{0} $ values is preferred. Consistent with these large values of $ M_{1/2} $, however, the
precision observable, $BR( b \rightarrow s+\gamma) \sim (3.36 \pm 0.23) \times 10^{-4}$ still allows one to constrict $ A_{0} $. The contrast of these indirect predictions for
$ M_{1/2} $ and $ A_{0} $ with the information from the unambigous detection of supersymmetric particles would provide a rigorous probing of the mSUGRA/CMSSM framework at the loop level.

\begin{center}
\begin{figure*}[htbp]
\centering{

\begin{subfigure}[]{\includegraphics[height= 6.9cm,width=8cm]{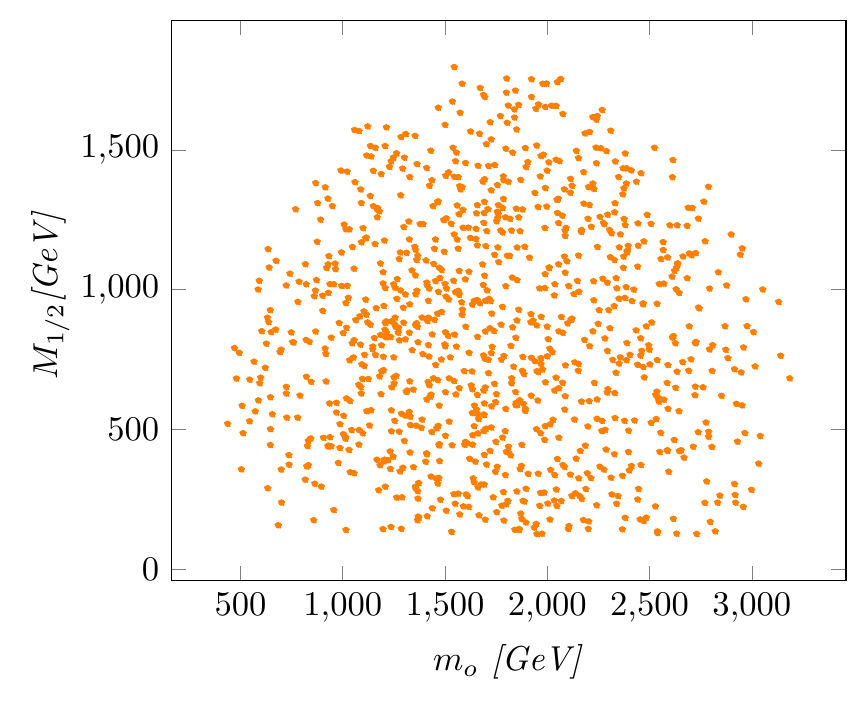}}\end{subfigure}
\begin{subfigure}[]{\includegraphics[height= 6.9cm,width=8cm]{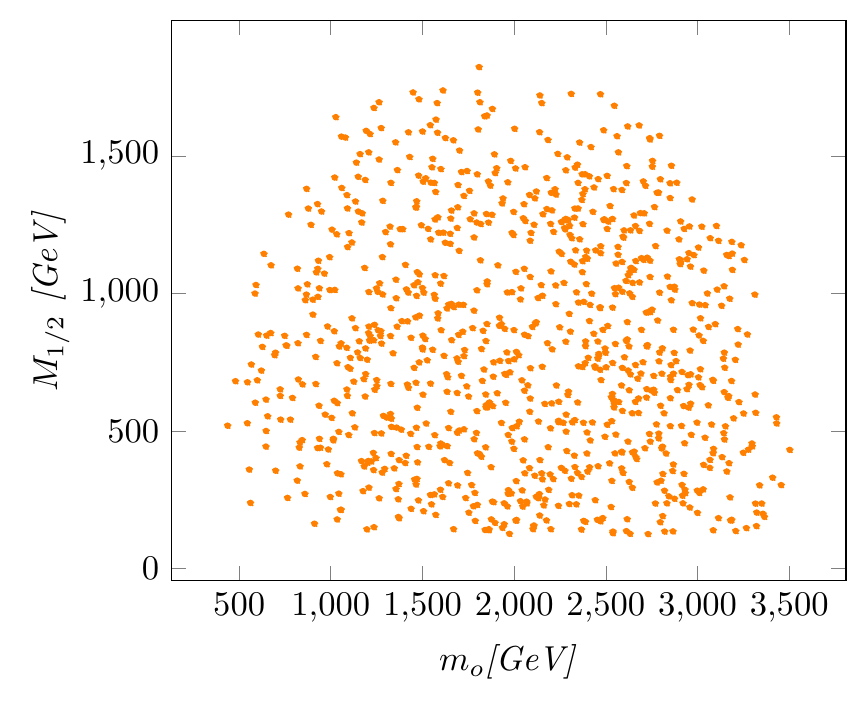}}\end{subfigure}\\

%\subfloat[]{\includegraphics[width = 3.5in]{3a.eps}} 
%\subfloat[]{\includegraphics[width = 3.6in]{3b.eps}}
 \caption{The CMSSM predictions for $m_{0}$ as a function of $ M_{1/2} $ tan $ \beta = 50 $ constrained by recent experimental bounds on BR($ b \rightarrow s+
\gamma$)  for various choices of $ A_{0} $. Figure in the left panel is for $ A_{0} \in \left[ -2M_{1/2} , + 2M_{1/2} \right]$ , and the figure in the right panel consider $ A_{0} $ to be lying in the range, $ A_{0} \in \left[ - 3 M_{1/2} , + 3 M_{1/2} \right]$}}
\end{figure*} 
\end{center}
\begin{center}
\begin{figure*}[htbp]
\centering{
\begin{subfigure}[]{\includegraphics[height= 6.9cm,width=8cm]{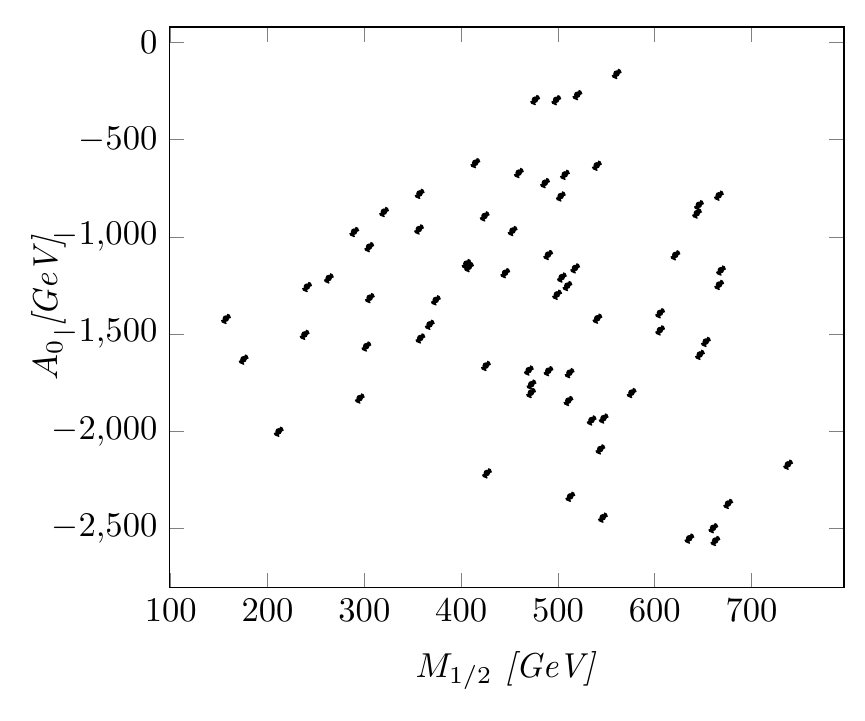}}\end{subfigure}
\begin{subfigure}[]{\includegraphics[height= 6.9cm,width=8cm]{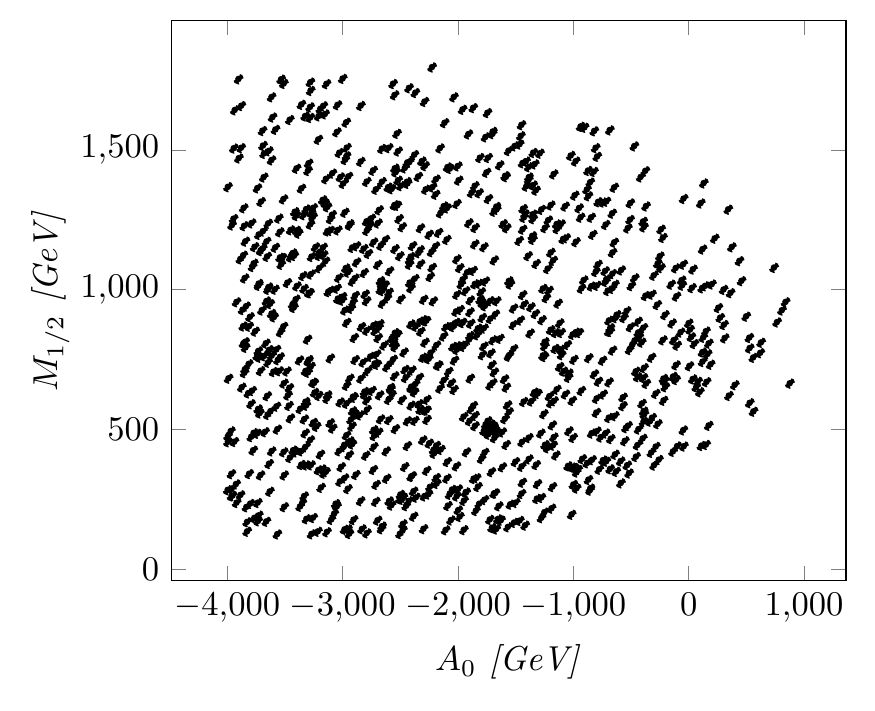}}\end{subfigure}\\
\begin{subfigure}[]{\includegraphics[height= 6.9cm,width=8cm]{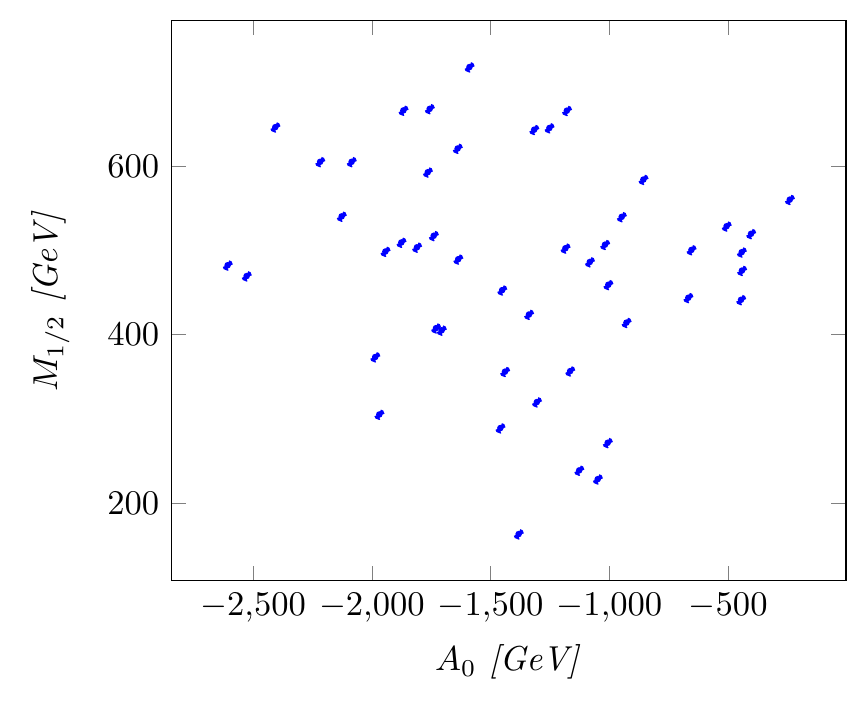}}\end{subfigure}
\begin{subfigure}[]{\includegraphics[height= 6.9cm,width=8cm]{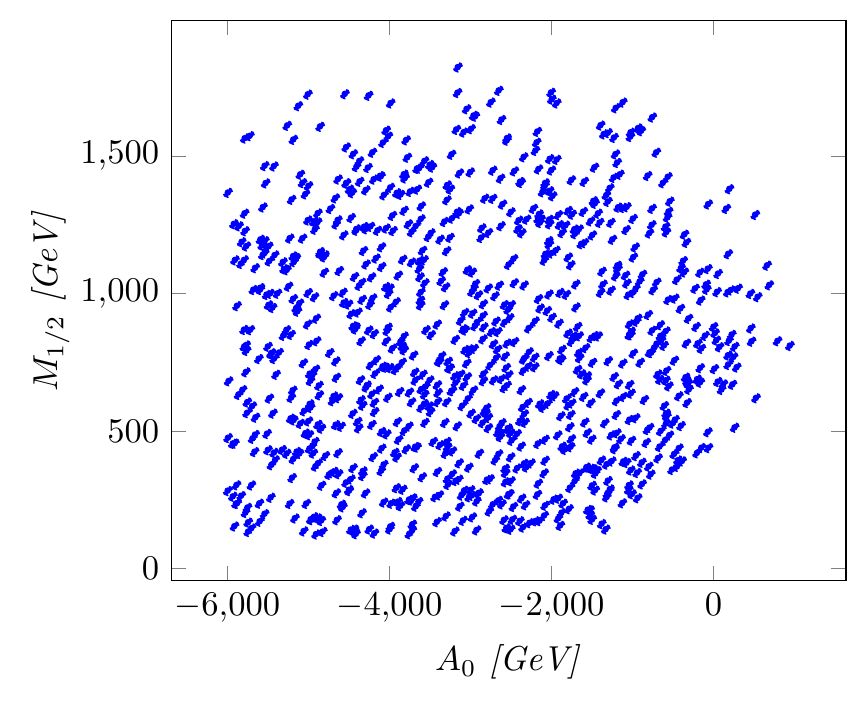}}\end{subfigure}\\
%\subfloat[]{\includegraphics[width = 3.5in]{3a.eps}} 
%\subfloat[]{\includegraphics[width = 3.6in]{3b.eps}}
 \caption{he CMSSM predictions for $ A_{0} $ as a function of $ M_{1/2} $ constrained by recent experimental bounds on BR($ b \rightarrow s+
\gamma$) for (a) tan $ \beta = 10 $ in the left panel and (b) tan $ \beta = 50 $ in the right panel and for various choices of $ A_{0} $. Figures in the top panel is for $ A_{0} \in \left[ -2M_{1/2} , + 2M_{1/2} \right]$, and the figures in the bottom panel consider $ A_{0} $ to be lying in the range, $ A_{0} \in \left[ -3M_{1/2} , + 3M_{1/2} \right]$.}}
\end{figure*} 
\end{center}

\par 
In this work in the CMSSM/mSUGRA model, the present experimental limit on $BR(b \rightarrow s \gamma)$ disfavors the soft SUSY breaking parameters $ M_{1/2} \leq $ 1 TeV  for $ m_{h} \sim$ 125.09 GeV. The flavor violating constraint
on the SUSY spectrum is relaxed if $ tan\beta = 50 $ is considered in mSUGRA model. 
\par 
In Tables {\color{blue}I}, a comparative study of the analysis in this work for the two cases, $ tan\beta = 50 $ and $ tan\beta = 10 $ is highlighted. The new results in CMSSM which one find in this work are the following:
\newline
1. Lighter $m_{0} \sim 500 $ GeV is allowed in mSUGRA.
\newline
2. A wider SUSY parameter space is allowed for $ tan \beta = 50$ as compared to $ tan\beta = 10$ .
\newline
3. Negative values of $A_{0}$ is preferred over positive values of $ A_{0} $ for the allowed susy parameter space.
\newline
4. BR($\mu\rightarrow e \gamma$) both increases and decreases with increase of scalar masses for $ A_{0}  = 0\hspace{.1cm} \text{TeV} $.

\section{\textbf{Conclusion}}
Here in this work the sensitivity of precision observable, BR($b \rightarrow s \gamma$), at LHC to detect the indirect effects of supersymmetry within the mSUGRA model is probed. The constraint BR($b \rightarrow s \gamma$) conclusively impoverish the
dimensionality of the CMSSM parameter space. Based on the present experimental results of the observable, BR($b \rightarrow s \gamma$), for two values of $ tan \beta $ taking into narration, the rare decay of b hadrons, $ b \rightarrow s\gamma $ in SUSY SO(10) theory in CMSSM/mSUGRA model, using the type I seesaw mechanism, is studied. 
The value of the Higgs mass compatible with the Standard Model (SM) Higgs boson in consistent with the ATLAS/CMS collaboration, the latest global data on the reactor mixing angle $ \theta_{13} $ for neutrinos, and the latest constraints on BR($ b  \rightarrow s  \gamma $) as predicted at the NNLO level for $ E_{\gamma} = 1.6$ GeV is used in this work. It is found that in mSUGRA, a light $ M_{1/2} $ region around 500 GeV $ - $ 700 GeV is allowed by the recent bounds of BR($ b  \rightarrow s  \gamma $) for $tan\beta = 10$, though in the $tan\beta = 50$ case a low $ M_{1/2} $ region starting from  around 250 GeV to values of  $ M_{1/2} $ as high as 1.5 TeV is also permitted. In mSUGRA, $ m_{0} $ the squark mass eigen values for  $tan\beta = 50$  as allowed by BR($ b  \rightarrow s  \gamma $) bound, shift toward a heavier spectrum around 3.5 TeV, as compared to   $tan\beta = 10$ where the squark mass eigen values $ m_{0} $ lies around 3 TeV. Thus in CMSSM, the allowed $ M_{1/2} $ parameter space at low energies becomes constrained. For a Higgs boson mass around 125.09 GeV, $ M_{1/2} $ $\leq $ 1.5 TeV is mostly disfavored for $tan\beta = 50$ and for $ A_{0} \in \left[ -2M_{1/2} , + 2M_{1/2} \right]$. Comparing these indirect predictions with the results from the direct scrutiny of supersymmetric particles will authorize a rigorous consistency test of this model at the loop level.
\par
The upshots presented in this work can effect the experimental indication for the generation of SUSY particles and can provoke a special detector set up to promise that
the substantial possible class of supersymmetric models steer to evident signatures at the future run of LHC. Consequently any study of heavy sparticles at the next run of LHC could assist us to percieve the nature of SUSY particles, in reference to curb put by the rare decay of b Hadrons. This in turn will lead towards a preferable understanding of theories beyond the standard model. 
\begin{table*}
\caption{Masses in this table are comparison between tan$ \beta =10 $, tan$ \beta=50 $ in this work for CMSSM.}
\label{sphericcase}
\begin{tabular*}{\textwidth}{@{\extracolsep{\fill}}lrrrrl@{}}
\hline\noalign{\smallskip}
\hline
Range of parameters allowed by & Range of parameters allowed by
\\  for BR $\left( b \rightarrow s \gamma\right), tan \beta = 10$  & BR $\left(b \rightarrow s\gamma\right), tan \beta = 50$  
\\ 
 \hline
 \textbf{1}.Figure 1a: & \textbf{1}. Figure 1b: $M_{1/2}\geq$ 0.5 TeV  to 1.5 TeV is allowed 
\\ Almost no $M_{1/2}$ space is allowed.\\ 
\textbf{2}.Figure 1c:& \textbf{2}. Figure 1d: A wider space of 
\\$M_{1/2}\sim$ 450 GeV to 650 GeV &  $ M_{1/2} $ from 250 GeV to 2 TeV is allowed.   
 &  
\\is allowed in this work. \\
\textbf{3}.Figure 2a,c:& \textbf{3}.Figure 2b,d:$M_{1/2}\sim$ 0.2 TeV 
\\$M_{1/2}\sim$ 0.5 TeV for & to 2 TeV is allowed.
\\to 0.6 TeV is permitted\\ 
\textbf{4}.Figure 3a,c,e: & \textbf{4}.Figure 3b: for $ m_{h} = 125.09$ GeV, no $ M_{1/2} $ space is allowed\\
 for $ m_{h} = 125.09$ GeV,& Figure 3e: for $ m_{h} = 125.09$ GeV, $ M_{1/2} \leq$ 1.5 TeV is disfavoured.\\
 no $ M_{1/2} $ space & Figure 3f: for $ m_{h} = 125.09$ GeV, $ M_{1/2} \leq$ 750 GeV is disfavoured.   \\
from 0 to 2 TeV is favoured. \\
\hline
\end{tabular*}
\end{table*}

\section*{Acknowledgments}
Gayatri Ghosh would like to thank Sudhir Vempati for fruitful discussions and suggestions of this problem at IISc Bangalore. She also thanks Barak Valley Engineering College, India where a major part of this work has been done. She would also like to thank AICTE India, for providing financial support to her.
 
%\begin{thebibliography}{000} %for 3 digits
%\begin{thebibliography}{00}  %for 2 digits

\end{document}